\documentclass{elsart1p}
 \usepackage{graphicx}

\usepackage{amssymb}

\begin{document}

\begin{frontmatter}



\title{Masses and semileptonic decays of doubly heavy baryons in a nonrelativistic quark model}


\author[1]{C. Albertus,}
\author[2]{, E. Hern\'andez}
\author[3]{, J. Nieves}
\author[2]{, and J. M. Verde-Velasco}
\address[1]{School of Physics and Astronomy, University of Southampton, Southampton 
SO17 1BJ, U.K.}
\address[2]{Departamento de F\'{\i}sica Fundamental, Universidad de Salamanca, E37008 Salamanca, Spain}
\address[3]{Departamento de F\'{\i}sica At\'omica, Molecular y Nuclear, Universidad de Granada, E18071 Granada, Spain}
\begin{abstract}
We evaluate masses and semileptonic decay widths for the ground state of doubly heavy $\Xi$ and $\Omega$ baryons in the framework of a
nonrelativistic quark model. We solve the three-body problem by means of a variational ansatz made possible by
heavy-quark spin symmetry constraints. Our masses are comparable to the ones obtained in relativistic calculations and
we get  one of the best agreements with lattice data. Our simple wave functions are used to evaluate semileptonic 
decays of doubly
heavy $\Xi,\,\Xi'(J=1/2)$ and $\Omega,\,\Omega'(J=1/2)$ baryons. Our results for the decay widths are in reasonable
agreement with calculations done in a relativistic calculation in the quark-diquark approximation. We also check that our wave 
functions comply with what it is expected in the infinite heavy quark mass limit.\vspace*{-.25cm}
\end{abstract}
\begin{keyword}
Nonrelativistic quark model \sep Hadron mass models \sep semileptonic decays\sep charmed and bottom baryons
\PACS 12.39.Jh \sep 12.40.Yx \sep 13.30.Ce  \sep 14.20.Lq \sep 14.20.Mr
\end{keyword}
\end{frontmatter}

\section{Introduction}
In this contribution we present results for masses and semileptonic decay widths of doubly heavy $\Xi$ and 
$\Omega$ baryons. The calculation has been done within a nonrelativistic quark model approach.
In hadrons with two heavy quarks, heavy quark symmetry manifest itself as a spin symmetry
 amounting to the decoupling of the heavy quark spins in the infinite
heavy quark mass limit~\cite{jenkins93}. In that limit one can consider the total spin
of the two heavy quark  subsystem  to be well defined.  In this
work we shall assume this is a good approximation for the actual heavy
quark masses that we use. This   approximation  
simplifies the solution of the baryon three-quark problem allowing for a 
simple variational ansatz for the orbital wave function. This approach leads to
simple and manageable wave functions and it was already applied, with obvious
changes,  in the study of baryons
with one heavy quark~\cite{conrado04}. In order to estimate part of 
the theoretical uncertainties
affecting  our calculation we have considered several simple
phenomenological quark--quark interactions taken from Refs.~\cite{BD81,SS94,Si96}. Their
free parameters have been adjusted in the meson sector and are thus
free of three--body ambiguities.  Due to  lack of space we shall concentrate 
in what follows 
in showing part of our results. Interested readers can find all details of the calculation 
and further results  in Ref.~\cite{epja07}.

\section{Some results}
In Table~\ref{tab:mass} we show the masses of doubly $cc$ and $bb$ baryons that have also been evaluated
in lattice QCD~\cite{khan00,lewis01,flynn03}, and the one experimental result for $M_{\Xi_{cc}}$
by the SELEX Collaboration~\cite{mattson02}. Our central results have been obtained with the AL1 potential of
Refs.~\cite{SS94,Si96},
while the errors shown  account for the spread in the results stemming from the use of different interquark interactions.
For doubly $c$ baryons we get good agreement with lattice data
whereas for doubly $b$ baryons our masses are a 100 MeV lower. The only experimental result has not been confirmed by
 other Collaborations and has at present a one-star status.
\begin{table}[h!!!]
\begin{center}
\begin{tabular}{lccccc}
& This work\hspace{.5cm}&Exp.$^{*}$~\cite{mattson02}
&Latt.~\cite{khan00}\hspace{.5cm} &Latt.~\cite{lewis01}\hspace{.5cm} &Latt.~\cite{flynn03} \\\hline
$M_{\Xi_{cc}}$\hspace{.5cm}  &$3612^{+17}$&$3519\pm1$&&$3605\pm23$&$3549\pm 95$ \\   
$M_{\Xi^*_{cc}}$  &$3706^{+23}$ &  &&$3685\pm 23$&$3641\pm97$\\   
$M_{\Xi_{bb}}$  &$10197^{+10}_{-17}$&&$10314\pm47$\\   
$M_{\Xi^*_{bb}}$  &$10236^{+9}_{-17}$& &$10333\pm 55$ \\ \hline
$M_{\Omega_{cc}}$\hspace{.5cm}  &$3702^{+41}$  &&&$3733\pm9^{+7}_{-38}$&$3663\pm 97$\\   
$M_{\Omega^*_{cc}}$ &$3783^{+22}$ &&&$3801\pm9^{+3}_{-34}$&$3734\pm 98$ \\   
$M_{\Omega_{bb}}$  &$10260^{+14}_{-34}$ &&$10365\pm40^{-11}_{+12}$$^{+16}_{-0}$\\   
$M_{\Omega^*_{bb}}$ &$10297^{+5}_{-28}$ &&$10383\pm39^{-8}_{+8}$$^{+12}_{-0}$\\ \hline
$M_{\Xi^*_{cc}}-M_{\Xi_{cc}}$&$94^{+5}_{-11}$&&&$80\pm11$ &$87\pm 19$ \\ 
$M_{\Xi^*_{bb}}-M_{\Xi_{bb}}$&$39^{+1}_{-6}$&&$20\pm6$ \\ 
$M_{\Omega^*_{cc}}-M_{\Omega_{cc}}$&$81^{+11}_{-19}$&&&$68\pm7$ &$67\pm 16$ \\   
$M_{\Omega^*_{bb}}-M_{\Omega_{bb}}$&$37^{+6}_{-9}$&&&$20\pm5$  \\ \hline  
\end{tabular}
\end{center}
\caption{Baryon masses and mass differences obtained in this calculation as compared with lattice data.}
\label{tab:mass}
\end{table}
Lattice simulations can also produce  independent determination of mass differences. Our mass differences are always
larger. The best agreement is reached for the potential in Ref.~\cite{BD81} for which we get always the lowest results.
Mass and mass differences comparisons with other models (different versions of the relativistic quark model, and lattice nonrelativistic QCD) can be found in Ref~\cite{epja07}.

With our simple wave functions we have further evaluated form factors, differential decay widths  and integrated 
decay widths for $b\to c$ driven semileptonic
transitions between doubly heavy baryons with total spin $J=1/2$. In Table~\ref{tab:dw} we show our results 
 for the  decay widths. We compare them with the results by Ebert {\it et al.}~\cite{ebert04} evaluated in a relativistic
 quark model in the quark diquark approximation and with the results by Guo {\it et al.}~\cite{guo98} obtained 
 with the use of the 
 Bethe-Salpeter equation applied to a quark-diquark system. We find a reasonable agreement with the calculation
 by Ebert {\it et al.} while the results in the Bethe-Salpeter approach of Guo {\it et al.} are much larger.
\begin{table}
\begin{center}
\begin{tabular}{lccccc}
  &This work\hspace{.25cm} &\cite{ebert04}&RTQM&\cite{guo98}&HQET\\ \hline
$\Gamma(\Xi_{bb}\to\Xi_{bc}\,l\bar\nu_l)$\hspace{.25cm}  &$ 3.84^{+0.49}_{-0.10}$&
3.26&&28.5&\\ 
$\Gamma(\Xi_{bc}\to\Xi_{cc}\,l\bar\nu_l)$  &$ 5.13^{+0.51}_{-0.05} $ & 4.59&0.79&8.93&4.0\\ 
$\Gamma(\Xi_{bb}\to\Xi_{bc}'\,l\bar\nu_l)$ &  $2.12 ^{+0.26}_{-0.05}$ &1.64&&4.28&\\ 
$\Gamma(\Xi_{bc}'\to\Xi_{cc}\,l\bar\nu_l)$ &  $2.71^{+0.19}_{-0.05}$ &
1.76&&7.76&\\ \hline
$\Gamma(\Omega_{bb}\to\Omega_{bc}\,l\bar\nu_l)$\hspace{.25cm}  &
$4.28^{+0.39}_{-0.03}$& 3.40&&28.8\\ 
$\Gamma(\Omega_{bc}\to\Omega_{cc}\,l\bar\nu_l)$ &$5.17^{+0.39}$&4.95&\\ 
$\Gamma(\Omega_{bb}\to\Omega_{bc}'\,l\bar\nu_l)$ &$2.32^{+0.26}$& 1.66&\\ 
$\Gamma(\Omega_{bc}'\to\Omega_{cc}\,l\bar\nu_l)$ &$2.71^{+0.17} $ &1.90&
\end{tabular}
\caption{Semileptonic decay widths in units of $10^{-14}$\,GeV. We have used a value
$|V_{cb}|=0.0413$. $l$ stands for a light charged lepton, $l=e,\,\mu$}
\end{center}
\label{tab:dw}
\end{table}

Finally comment that we have also checked that our  variational wave functions have the correct infinite heavy quark mass limit.
 In that limit the wave function  should look like the one for a meson composed of a light quark and a heavy
 pointlike diquark. In our model the pointlike nature of the heavy diquark comes about through the one-gluon 
 exchange Coulomb  potential, present in the interquark interactions we use, that binds the two heavy quark into a distance given by
 the inverse of their reduced mass. Besides we have shown that as the heavy quark mass increases our wave function
 reduces to the product of the ground state wave function for the heavy diquark subsystem times the ground state wave
 function for the relative motion of the light quark with respect to the center of mass of the heavy diquark.\\

{\small
 This research was supported by DGI and FEDER funds, under contracts
FIS2005-00810, BFM2003-00856, FPA2004-05616 and FIS2006-03438, by Junta de
Andaluc\'\i a and Junta de Castilla y Le\'on under contracts FQM0225,
SA104/04 and SA016A07, and it is part of the EU integrated infrastructure
initiative Hadron Physics Project under contract number
RII3-CT-2004-506078.  J. M. V.-V. acknowledges an E.P.I.F. contract  with the
University of Salamanca.}

\end{document}